# On Mechanical Behavior of Elastomeric Networks: Effects of Random Porous Microstructure

Mohammad Tehrani[1], Mohammad Hossein Moshaei[1], Mehdi Jafari[2], Mohammad Khalili[2, *]

*Ohio University, Athens, Ohio, 45701[1]*
*Khomeinishahr Branch, Islamic Azad University, Khomeinishahr, Iran[2]*

Corresponding author[*]
*Email address:* dr.mo.khalili@gmail.com (Mohammad Khalili)

**Abstract**
An assumption in micromechanical analysis of polymers is that the constitutive polymeric media is non-porous. Non-porosity of media, however, is merely a simplifying assumption. In this paper, we neglect this assumption and studied polymer networks with a different porosity volume fraction. A random morphology description function is used to model the porosity of the network and nonlinear finite element analyses are conducted to perform structural analysis of porous polymer networks. The results show that the porosity effect is significant in mechanical behavior of polymer networks and may increase the maximum Von-Mises stress drastically.

1. **Introduction**

   Since the discovery of zeolites and their successful industrial applications, the porous materials have become one of the most exciting frontiers in modern science such as chemistry and physics. From physical point of view, porosity plays significant role in modifying properties of the materials and has variety of applications including medical devices and tissue engineering [1, 2]. In the past decade, the field of porous materials has undergone revolutionary growth. A number of new porous materials such as metal-organic frameworks (MOFs), crystalline covalent-organic frameworks (COFs), and amorphous porous organic polymers (POPs) have been well developed and intrigued much attention not only owing to their high porosity like conventional porous materials such as zeolites or activated carbons, but also their capable of incorporating targeted or multiple chemical functionalities into the porous framework by bottom-up or post-synthetic modification approach. They have been recently explored as promising candidates for applications in gas storage and gas separations [3]. These applications of porous polymers make it essential to perform fluid-structure interaction (FSI) analysis in presence of porous polymers as the solid structure. While there has been conducted numerous studies over numerical analysis [4] and CFD methods and its applications [5, 6, 7, 8, 9, 10, 11, 30], investigating the mechanical properties and finite element analysis of porous polymers remain intact. The objective of this paper is to develop and implement a methodology for creating morphologically realistic heterogeneous random porous microstructures over the entire volume fraction range, and subsequently analyze their statistical and homogenized material properties in an effort to extract valuable insight into the behavior of realistic porous polymer. To that end, the asymptotic expansion homogenization (AEH) method is used in conjunction with multiscale analysis to obtain the stresses at the microscopic and macroscopic levels.

   **2. Finite Element Model (FEM)**

The random morphology description function (RMDF) is implemented to build random microstructure models with different porosity values. Where N is number of random function, $C_i$ and $y^i$ are random coefficient and random coordinates respectively.

$$f(y) = \sum_{i=1}^{N} c_i e^{-[\frac{(y_1-y_1^{(i)})+(y_2-y_2^{(i)})}{w_i^2}]}, \quad w_i = \frac{1}{\sqrt{N}} \tag{1}$$

This equation is a summation of two-dimensional Gaussian functions to create realistic random function [12]. Using Eq. (1) with different values of cutoffs used to build 2D random porous media with different amount of porosity.

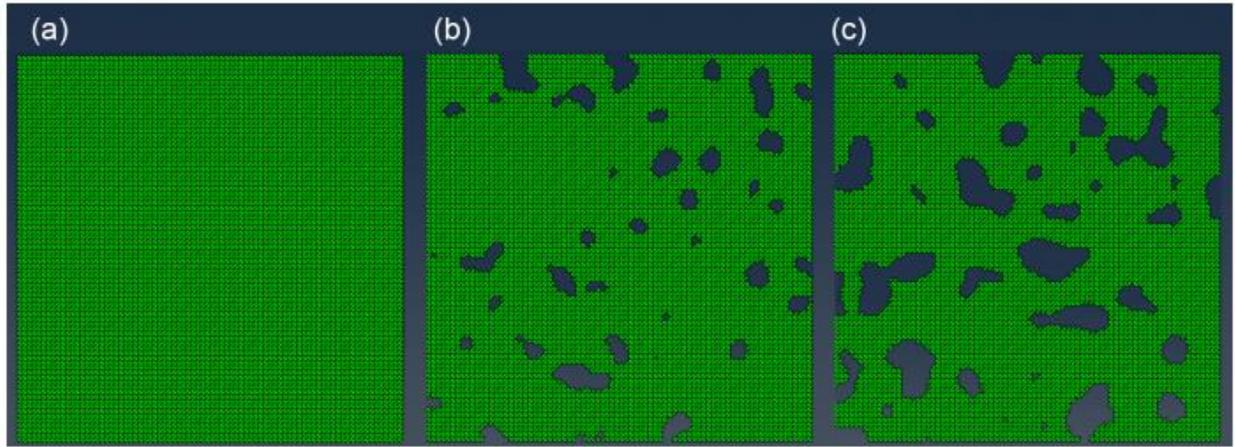

Figure 1: (a) nonporous media, (b) porous media with 10% porosity, and (c) porous media with 20% porosity.

A couple of phenomenological-based models and mechanical statistics-based models [13, 14] has been developed to model mechanical response of the polymer networks. Recently, Tehrani-Sarvestani model [15, 16, 17, 28] has been developed to model the mechanical behavior and failure of polymer networks. In this study, the Neo-Hookean model is used. For a compressible neo-Hookean material the strain energy density function is given by,

$$W = C_1(\bar{I}_1 - 3) + D_1(J-1)^2 \tag{2}$$

where $D_1$ and $C_1 D_1$ are material constants, $\bar{I}_1 = J^{-2/3} I_1$ is the first invariant of the isochoric part of the right Cauchy-Green deformation tensor. Table 1 indicated the mechanical properties of the polymer network in this simulation. The size of the specimen is 100*mm*× 100*mm* and 6-node modified quadratic plane stress triangle elements (CPS6M) were used for the model in order to reduce mesh density without affecting solution accuracy. Figure 5, shows the schematic boundary condition of the finite element model. The implemented model is subject to a biaxial deformation control loading.

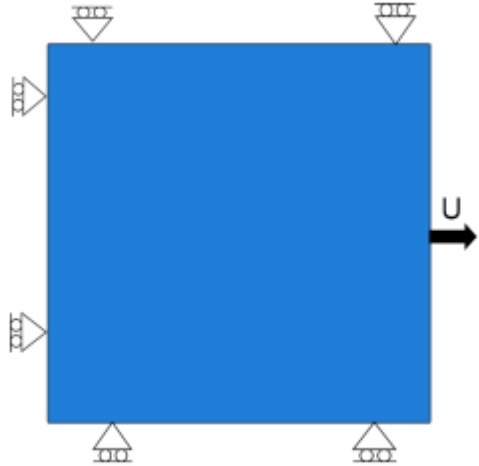

Figure 2: Schematic boundary condition of finite element model.

Table 1: Mechanical property of polymer network [18].

| Parameter | Value |
|---|---|
| $C_1$ (MPa) | 0.2587 |
| $D_1$ | $1.5828e^{-3}$ |

## 3. Results and Conclusion

In this section, the mechanical behavior of homogenized polymer networks is presented for heterogeneous random porous polymer networks. Von-Mises stresses and total deformations of the models are illustrated in Figure 6-7. While the maximum Von-Mises stress in the non-porous media is found as 1.9 (MPa), the maximum Von-Mises stress in porous media with $v$ = 10% and $v$ = 20% is 21.9(MPa) and 22.6(MPa) respectively. The stress contours in the porous media indicate that by increasing the porosity of the media, the maximum Von-Mises stress will increase drastically.

In comparison with non-porous media, by increasing the porosity of the media, the higher volume fraction of the porous media will face the same or lesser amount of the Von- Mises stress in the non-porous media-i.e. lesser volume fraction of the porous media will withstand higher values of stresses-

It can be observed from the stress contours that the certain regions, e.g. between the holes, will face extremely higher values of stresses due to the geometry of porous media and these regions are more prone to failure.

In attempt to correlate the mechanical behavior of porous polymer network and porosity fraction, a finite element framework is established. Randomly and heterogeneously are main assumptions of this framework. We have shown that structural analysis using FE method of porous media is crucial to investigate. In this manner, finite element method has been successfully used to develop mechanical properties of the heterogeneous random porous polymer network. Initial simulations have shown that increasing the porosity will increase the maximum Von-Mises stress concurrently. This study shows that in presence of porous polymer networks, performing structural analysis to predict the stress field and failure is inevitable. While there are many ongoing studies in the field of nanoparticles [19-25], this study illustrates the importance of more investigation for polymers with nano-porosity.

There are remarks regarding capability and validity of proposed framework which must be mentioned. First, the polymer network is assumed to be polydisperse and chains length distribution follows simple exponential distribution [17, 26, 27, 29]. Second, the deformations are assumed to be affine which means that at each instance all the polymer strands are under the same deformation. [15, 16]

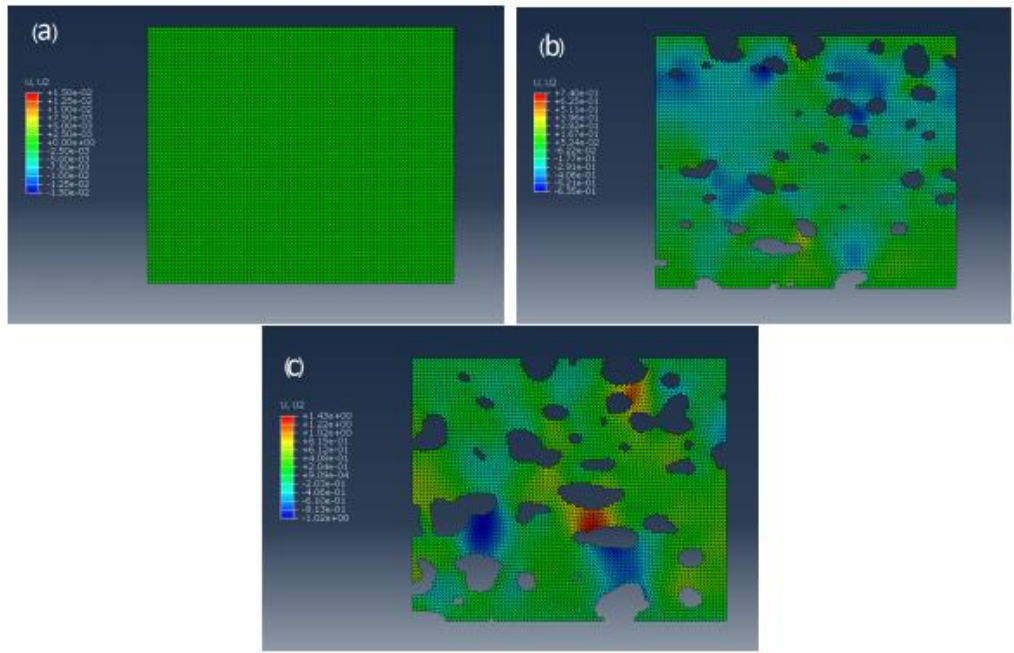

Figure 3: deformation perpendicular to the direction of subjected load with (a) no porosity (b) 10% porosity (c) 20% porosity.

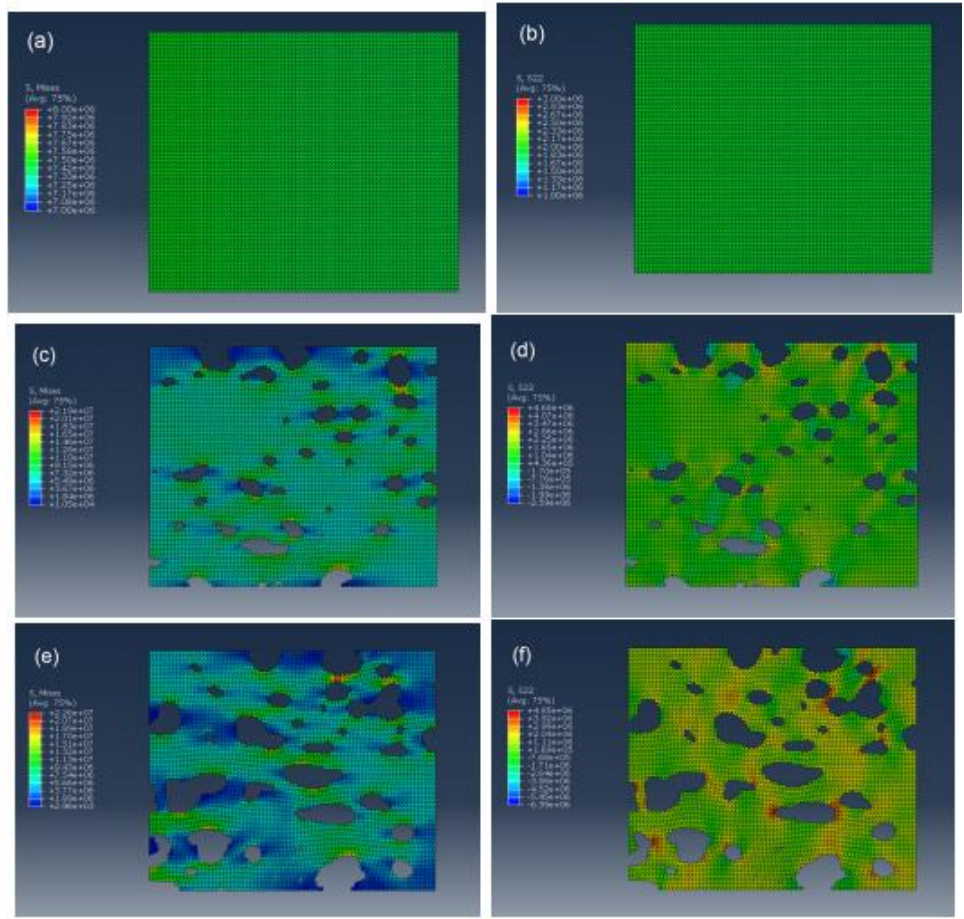

Figure 4: Vin-Mises stress of media with (a) no porosity (c) 10% porosity (e) 20% porosity and stress perpendicular to the direction of subjected load with (b) no porosity (d) 10% porosity (f) 20% porosity